\documentclass[prd,twocolumn,showpacs,amsmath,amssymb]{revtex4}
\usepackage{graphicx}
\begin{document}
\title{\textbf{Cosmology with extended Chaplygin gas media and $w=-1$ crossing}}                                
\author{Xin-He Meng}
\email{xhm@nankai.edu.cn}
\author{Ming-Guang Hu}
\email{hu_mingguang@hotmail.com}
\author{Jie Ren}
\email{jrenphysics@hotmail.com}
\affiliation{Department of
physics, Nankai University, Tianjin 300071, China}
\begin{abstract}                                                
In this paper the accelerating
expansion of our universe at the late cosmic evolution time in a generally
modified (extended) \emph{Chaplygin gas} (Dark Fluid) model is detailed
, which is characterized by two parameters ($m$, $\alpha$).
Different choices to the parameters $m$ and $\alpha$ divide this model
into two main kinds of situation by different properties with
cosmological interests. With proper choices of parameters, we find
that this extended model can realize the phantom divide $w=-1$
(Equation Of State parameter) crossing phenomenon with interesting ranges
of the scale factor value $a(t)$ corresponding to $w=-1$ and present
value of state parameter $w$.

Additionally, through Taylor series expansion of the function
$A(a)$ in the extended Chaplygin gas  model
,
a specific Equation of State is gained.
Under the framework of Friedman-Robertson-Walker cosmic model, it
can successfully explain the accelerating expansion of our universe.
However, the value of  $w$ in this case is large than $-1$, that
is, indicating it like a quintessence fluid and no $w=-1$ crossing
occurs.

\end{abstract}
\pacs{98.80.Cq, 98.80.-k} \maketitle

\section{Introduction}                                         
Dark side of the Universe has been puzzling us across the century \cite{sc},
especially the recent years discovery as coined Dark Energy.
Observations of type Ia supernova(SNe Ia) directly suggest that
the expansion of the universe is accelerating, and the measurement
of the cosmic microwave background (CMB) \cite{DNS} and the galaxy
power spectrum for large scale structure\cite{MT} indicate that in
spatially flat isotropic universe, about two-thirds of the
critical energy density seems to be stored in a so called dark
energy component with enough negative pressure responsible for the
currently cosmic accelerating expansion\cite{AGR}. It is clear
from observations that most of the matter in the Universe is in a
dark (non-baryonic) form (see, for instance, \cite{pd}). To
understand the mysterious dark components is even great challenge for
temporary physicists from this new century crossings.

 The simplest candidate for the dark energy is a
cosmological constant $\Lambda$, which has a specially simple
pressure expression $p_{\Lambda}=-\rho_{\Lambda}$. However, the
$\Lambda$-term requires that the vacuum energy density be fine tuned
to have the observed very tiny value, the famous "old" cosmological
constant problem. So, many other different forms of dynamically
changing dark energy models have been proposed instead of the only
cosmological constant incorporated model. Usually, the equation of
state(EOS) for describing dark energy can be assumedly factorized into the
form of $p_{DE}=w\rho_{DE}$, where $w$ may depend on cosmological
redshift $z$ \cite{DP}, or scale factor $a(t)$ and so on. By the
way, the case for $w=-1$ corresponding to the cosmological constant,
which has involved singularity in perturbation calculations, was
thought as border-case called the phantom divide\cite{WH}.

To further explore the properties of the so called dark energy or
its concrete form of EOS $w$, some authors propose kinds of models
such as $p_{X}=-\frac{A}{\rho_{X}}$(Chaplygin gas model\cite{AK}),
which is not consistent with astrophysics observations as it can
produce oscillations and exponential blow-up to the matter power
spectrum, its generalizations \cite{SN} like $p_{X}=-\frac{A}{\rho_{X}^{\alpha}}$(generalized Chaplygin
gas model(GCGM)\cite{MCB, GD}), or generalizing the constant A to variable
forms\cite{ZKG}, $p_{X}=\alpha(\rho_{X}-\rho_{0})$(linearized EOS
model\cite{EB}, compared with the conventionally perfect fluid EOS
$p_{X}=(\gamma-1)\rho_{X}$), and so on(subscript $X$ represents dark
energy fluid except for the cosmological constant). Among these
models, the fluid of GCGM has a dual behavior: it mimics
matter($p\approx0$) domination at very early stage in the history of
the universe evolution and a cosmological constant much later(with a
smooth transition in between), which is highly suggestive of a
unified description of dark matter and dark energy (see reference
\cite{LMGB}) that we may address it as Dark Fluid, as reason above. Moreover, dark
energy may be made up of different energy fluids with different EOS,
and these different energy fluids interact with each other through
$\rho_{s}'=\frac{\rho_{s}+p_{s}}{\rho+p}=\frac{d\rho_{s}}{d\rho}$,
where subscript s indicates a component,  which is called the
 self-interacting fluids(see reference\cite{LPC}).

With the merits of GCGM, we further discuss a modified general
Chaplygin gas model by extending its EOS as an explicitly scale
factor related form, which is based on the original relation
$p_{X}=-\frac{A}{\rho_{X}}$. Moreover, through the mathematically
Taylor series expansion of the relatively general EOS we elaborately
demonstrate concrete values of two characteristic parameters ($m$,
$\alpha$) for this ECG model, and we find the cosmic effective compositions
(matter, radiation and vacuum energy) are from the extended Chaplygin gas component.

The plot of this paper is arranged as: In Sect.\ref{sec:modified
chaplygin gas}, first we introduce the framework by mathematically
integrable expression of the extended Chaplygin gas model with
analytical investigations. From the mathematical point of view, we
divide the model into two classes described by two characters $m$
and $\alpha$. And then in Subsects.IIA and
IIB, we investigate the two classes with general
theoretical discussions, respectively. Secondly, through the Taylor
series expansion method , a set of parameters $(m, \alpha)$ is
determined and its physical indications are presented in
Sect.\ref{sec:The Taylor expansion case}. At the last part, we
summarize with discussions.

\section{Extended Chaplygin gas model}                  \label{sec:modified chaplygin gas}
The metric of a homogeneous and isotropic universe is usually
written as follows
\begin{equation}
ds^{2}=dt^{2}-a^{2}(t)dl^{2},
\end{equation}
where $dl^{2}$ is the metric of a 3-manifold of constant curvature
$(k=0,\pm1)$, and $a(t)$ is the expansion factor.

Under the framework of Friedman-Robertson-Walker cosmology, 
the dynamic evolution of cosmology is completely determined by the
Friedman equation
\begin{equation} \label{eq:Fri}
\frac{\dot{a}^{2}}{a^{2}}+\frac{k}{a^{2}}=\rho
\end{equation},
energy conservation equation
\begin{equation}                    \label{eq:energy conservation}
d(\rho a^{3})=-pd(a^{3})
\end{equation}
and equation of state $p=f(\rho)$ if we have. With the
consideration that our Universe is filled mainly with both matter
(mostly non-relative dark matter) and dark energy( the radiation
component negligible at the later phase for the Universe
evolution), so in the above Eq.(\ref{eq:energy conservation})
$\rho=\rho_{m}+\rho_{DE},\quad p=p_{m}+p_{DE}$ and we have used
the unit convention $8\pi G/3=c=1$ throughout.

As for EOS, taking $p=f(\rho)$ into Eq.(\ref{eq:energy
conservation}), we get generally:
\begin{equation}
\int\frac{d\rho}{\rho+f(\rho)}=-3\int\frac{da}{a}
\end{equation}
If only we know the concrete form of the function $f(\rho)$,  the
evolution of scale factor $a$ with the energy density $\rho$, may
be deduced out.

 In order to describing the behavior of dark energy with more possibilities,
an extension of the \emph{Chaplygin gas} model
is considered, and its relatively general EOS takes the following
form:
\begin{equation}                                \label{eq:state equation}
p_{X}=-\frac{A(a)}{\rho_{X}^{\alpha}}\rho_{X}
\end{equation}
where $p_{X}$ is pressure, $\rho_{X}(positve)$ with $\alpha>0$ is
energy density of the dynamically changing part for the dark energy
component if we also include the cosmological constant contribution,
and $A(a)$ is a positive function depending on scale factor $a$ or
its inverse. In some generalized Chaplygin gas model the cosmological
constant can not be obtained at later cosmic times. The total EOS of
dark energy can be, therefore, divided into two parts---the
cosmological constant $\rho_{\Lambda}$ and dynamically changing part
$\rho_{X}$. That is formerly:
\begin{equation}
\rho_{DE}=\rho_{\Lambda}+\rho_{X},\qquad p_{DE}=p_{\Lambda}+p_{X}
\end{equation}
where the interrelations are,
\begin{equation}
p_{\Lambda}=-\rho_{\Lambda},\qquad
p_{X}=-\frac{A(a)}{\rho_{X}^{\alpha}}\rho_{X}
\end{equation}
Also Eq.(\ref{eq:energy conservation}) is divided into
, with the assumptions they decouple each other:
\begin{eqnarray}
d(\rho_{m}a^{3})=-p_{m}d(a^{3}),\qquad
d(\rho_{\Lambda}a^{3})=-p_{\Lambda}d(a^{3}),\nonumber
\end{eqnarray}
\begin{equation}
d(\rho_{X}a^{3})=-p_{X}d(a^{3})\label{eq:energy conservation beta}
\end{equation}
i.e, by assuming that there is no interaction between different
components. In this paper, our main attention is put on the
`$X$-component'.

 When $A(a)$ takes a constant value $A$ and $\alpha=2$,
 it results in the following relation, the CG case (see reference \cite{AK}):
 \begin{equation}
 \rho_{X}=\sqrt{A+\frac{B}{a^{6}}} \nonumber
 \end{equation}
where $B$ is an integrated constant. We can get a more general
relationship between density $\rho_{X}$ and scale factor $a$ by
considering(\ref{eq:state equation}) and (\ref{eq:energy
conservation beta}) , that is
\begin{equation}                            \label{eq:pp}
\int^{a_{0}^{3\alpha}\rho_{X,0}^{\alpha}}_{a^{3\alpha}\rho_{X}^{\alpha}}
d(a^{3\alpha}\rho^{\alpha})=\int^{a_{0}}_{a} 3\alpha
a^{3\alpha-1}A(a)da
\end{equation}
where $a_{0}$ is the present value of the scale factor and
$\rho_{X,0}$ is the present $X$-term energy density.

Now entering the essence of this paper, we consider the function
$A(a)$ with the form $A(a)=A_{0}a^{-m}$($A_{0}$ is constant,and
$m>0$), and then the characteristics of our model are determined by
three undetermined parameters $(m,\alpha,A_{0})$. A special case
with $\alpha=2$ results in the following relation\cite{ZKG}:
\begin{equation}
\rho_{X}=\sqrt{\frac{6}{6-m}\frac{A_{0}}{a^{m}}+\frac{B}{a^{6}}}\nonumber
\end{equation}
Obviously it can easily return to the original CG model result by
taking m=0 directly.

 Otherwise, in the point of view for integrability to (\ref{eq:pp}),
two cases are obvious as following. We will discuss $m=3\alpha$ in
the section \ref{sec:m=3n} and $m\neq3\alpha$ in the section
\ref{sec:m><3n}. We have known that present observation data
constrains the range of the equation of state parameter for dark
energy as $-1.38<w<-0.82$. Hao and Li \cite{HJG} have demonstrated
that $w=-1$ state is an attractor for the Chaplygin gas model and
the equation of state of this gas could approach this attractor from
either $w<-1$ or $w>-1$ sides. So in following sections, we would
also consider this attractor approaches on the background of the
extended \emph{Chaplygin gas} model.
\subsection{The $m=3\alpha$ case}         \label{sec:m=3n}             
In this section, we take $m=3\alpha$ so Eq.(\ref{eq:pp}) gives the
following result:
\begin{equation}  \label{eq:rho^n}
\rho_{X}^{\alpha}=3\alpha A_{0}\frac{\ln
\frac{a}{a_{0}}}{a^{3\alpha}}+\frac{a_{0}^{3\alpha}\rho_{X,0}^{\alpha}}{a^{3\alpha}}
\end{equation}
We can see that,
\begin{eqnarray}
&\rho_{X}\propto a^{-3}, \qquad \quad\quad a\rightarrow a_{0} \nonumber\\
&\rho_{X}\propto a^{-3}(\ln a)^{\frac{1}{\alpha}}, \quad
a\rightarrow\infty\nonumber
\end{eqnarray}
It means that around $a=a_{0}$ the dynamically changing part dark
energy density evolves similar to dust-like matter with $\rho\sim
a^{-3}$. In contrast, it evolves slower as $a\rightarrow\infty$
than that of $a\sim a_{0}$.

\begin{figure}
\begin{center}
\includegraphics{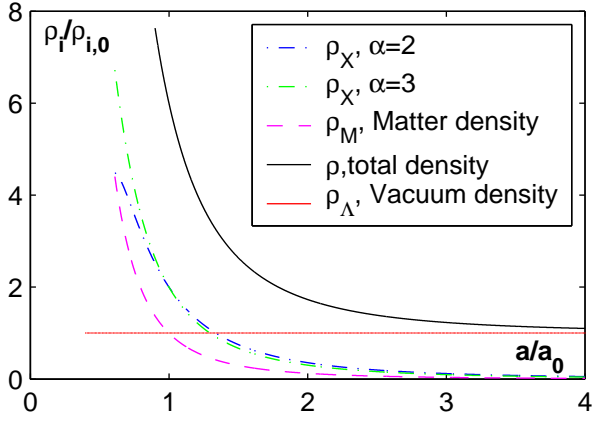}
\caption{On the $\rho_{i}-a$ plane, $\rho_{i}$ changes depending on
$a$. $i$ represents $M,X,\Lambda$.(assuming $A_{0}=1$). And The
total density $\rho=\sum_{i}\rho_{i}$ is also showed up. }
                            \label{mn13}
\end{center}
\end{figure}

\begin{figure}
\centering
\includegraphics{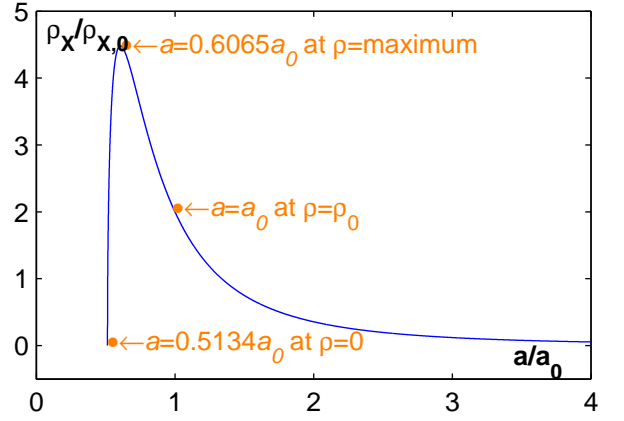}
\caption{On the $\rho_{X}-a$ plane, $\rho_{X}$ changes depending on
$a$. With $a(t)$ increasing, $\rho_{X}$ from $0$ increases to a
positive maximum and then decreases to approach a constant near
zero.} \label{mnpq}
\end{figure}

Figure \ref{mn13} shows

from $a=0.5134a_{0}$ the zero ECG density point, and with the
increasing of scale factor $a$, the dynamically changing part of
dark energy density has reached a biggest value (for example:
corresponding to $\alpha=2$ case,
$(\rho_{X})_{max}=\rho_{X}(0.6065a_{0})$ seeing Figure \ref{mnpq}),
and then it approaches to a constant near zero.

As to $\alpha=0$, we gain $p=-A_{0}\rho$, which can be described
directly by conventional EOS $p=w\rho$($w$ is constant). More
discussions for this situation can be found in references
\cite{EB,LMGB,MAM}, for example.

Taking Eq.(\ref{eq:rho^n}) into Fridamman Eq.(\ref{eq:Fri}), we get:

\begin{equation*}
\int^{a}_{a_{0}}\frac{a^{1/2}da}{(3\alpha A_{0}\ln
a/a_{0}+a_{0}^{3\alpha}\rho_{X,0}^{\alpha})^{1/2\alpha}}=\int^{t}_{t_{0}}
dt
\end{equation*}
\begin{figure}
  \includegraphics{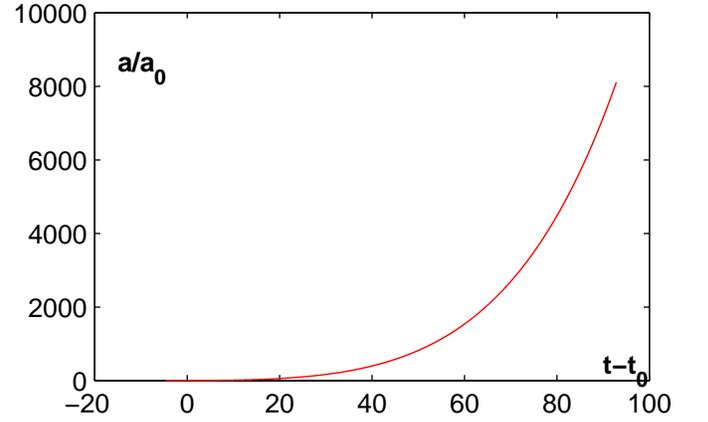}\\
  \caption{The evolution picture of $a(t)$ depends on $t$, and we take $\alpha=1/2$.}\label{fig:Charplygin1}
\end{figure}

Because the uncertainty of $\alpha$, the above integration is not
easy to solve. But for a general picture about the scale factor, we
can simply put $\alpha=1/2$. Thus we get the integration result:
\begin{equation}\label{eq:Yt}
    \ln
    Y+Y+\frac{Y^{2}}{2\cdot2!}+\frac{Y^{3}}{3\cdot3!}+\cdots=\frac{3A_{0}}{2a_{0}^{3/2}}e^{\sqrt{a_{0}^{3}\rho_{X,0}}/A_{0}}t+C
\end{equation}
where
$Y=\frac{3}{2}\ln\frac{a}{a_{0}}+\sqrt{a_{0}^{3}\rho_{X,0}}/A_{0}$.
Eq.(\ref{eq:Yt}) can be directly  figured as in \ref{fig:Charplygin1},
which explicitly shows scale factor in a accelerating state with
$\ddot{a}>0$ and there is no singularity during the process.

 Since no singularity(see reference
\cite{SN} for the classification of singularity) occurs during the
evolution of $\rho_{X}(a)$, we can discuss the $w=-1$ crossing with
it. So far as we know, $w=\frac{p_{DE}}{\rho_{DE}}$ results in the
following equations with subscript denoting today's value:
\begin{eqnarray}                            \label{eq:omega}
w&=&-\frac{\frac{\rho_{X}}{3\alpha\ln(a/a_{0})+1/\eta}+\rho_{\Lambda}}
{\rho_{X}+\rho_{\Lambda}}                   \nonumber  \\
w_{0}&=&-\frac{\eta\rho_{X,0}+\rho_{\Lambda}}
{\rho_{X,0}+\rho_{\Lambda}}
\end{eqnarray}
where $\eta$ is defined by $\eta=A_{0}/(a^{m}\rho^{\alpha}_{X})$
called Judge for its following property.

Taking (\ref{eq:rho^n})into above equation, we can get:
\begin{displaymath}
\ln\frac{a}{a_{0}}\left( \begin{array}{l}
<\\
=\\
>
\end{array} \right)
\frac{\eta-1}{3\alpha \eta}
\quad provided
 \left(
\begin{array}{l}
w<-1\\
w=-1\\
w>-1
\end{array}
\right )
\end{displaymath}
Then we can see that through the evolution of scale factor $a(t)$,
the value of state parameter $w$ changes continuously. So that
crossing state barrier $(w=-1)$ would occur during the procedure of
the evolution. It has also told that Judge $\eta$ here played an
important role of relating to the different ranges of the present
state parameter $w_{0}$ and the scale factor (denoted by $a^{w}$) in
accordance with Phantom Divide($w=-1$) crossing as tabled in
\ref{table:m=3n} below.

\begin{tabular*}{7cm}{@{\extracolsep {\fill}}c|c|c}
 \multicolumn{3}{c}{Table \ref{table:m=3n}. Parameter ranges and $w=-1$  }\\
  \hline
  $\eta$ & $w_{0}$& $a^{w} (at\quad w=-1)$ \\
  \hline
  $=1$ &$ =-1$ & $=a_{0}$\\
  \hline
  $<1$ & $>-1$ & $<a_{0}$ \\
  \hline
  $>1$ & $<-1$ & $>a_{0}$\\
  \hline

\end{tabular*}  \label{table:m=3n}

Why can the Judge $\eta$ have such important property? In the next
part, we will work out its physical meaning in the more complete
background.

Now, as for eq.(\ref{eq:Fri}), we can get its equivalent expression
as:
\begin{eqnarray}                            \label{eq:H2(m=3n)}
h^{2} &=&
      \frac{\rho_{m}}{H_{0}^{2}}+\frac{\rho_{X}}
      {H_{0}^{2}}+W_{\Lambda,0}-\frac{k}{H_{0}^{2}a^{2}} \nonumber\\
      &=&W_{m,0}(1+z)^{3}+W_{X,0}(1+z)^{3}\nonumber\\
      && [3\alpha\eta\ln(1+z)+1]
      ^{\frac{1}{\alpha}}+W_{\Lambda,0}\nonumber\\
      &&+W_{k,0}(1+z)^{2}
\end{eqnarray}
where $h(z)$ with $h(z)=\frac{H(z)}{H_{0}}$ is reduced Hubble
parameter and $z=a_{0}/a-1$ is red-shift,
$W_{m,0}=\frac{\rho_{m,0}}{\rho_{c,0}}$,
$W_{X,0}=\frac{\rho_{X,0}}{\rho_{c,0}}$,
$W_{\Lambda,0}=\frac{\Lambda}{3\rho_{c,0}}$, and
$W_{k,0}=\frac{k}{a_{0}^{2}\rho_{c,0}}$($\rho_{m,0}$, $\rho_{X,0}$,
$H_{0}$, and $a_{0}$ are, respectively, present non-relative matter
density, dynamically changing part dark energy density, Hubble
parameter, and scale factor); $\rho_{c,0}=H_{0}^{2}$ is the current
critical density. These density parameters should satisfy the
following relation:
\begin{equation}
W_{m,0}+W_{X,0}+W_{\Lambda,0}+W_{k,0}=1
\end{equation}

Previously, we have pointed out that at the stage of $a\rightarrow
a_{0}$ the density $\rho_{X}$ evolves like dust-like matter and now
we can also see this point from following. Around $a=a_{0}$, we can
expand the term $[3\alpha\eta\ln(1+z)+1]
      ^{1/\alpha}$ around $z=0$,  and preserve the first order of $z$
      and plus other terms. Thus, we get:
\begin{eqnarray}                            \label{eq:H20(m=3n)}
h^{2}&=&\widetilde{W}_{m,0}(1+z)^{3}+3W_{X,0}\eta(1+z)^{4}+\nonumber\\
      &&+W_{\Lambda,0}+W_{k,0}(1+z)^{2}
\end{eqnarray}
The equivalent density parameter $\widetilde{W}_{m,0}$ now is
defined by:
\begin{equation}
\widetilde{W}_{m,0}=W_{m,0}+W_{X,0}(1-3\eta)
\end{equation}
which has included the dark energy density contribution part. The
$(z+1)^{4}$ term corresponds to the radiation component, which
can be neglected in the beginning.

Considering the deceleration parameter $q=-a\ddot{a}/\dot{a}^{2}$
, it can be expressed by the use of reduced Hubble parameter as:
\begin{equation}                        \label{eq:q}
q(z)=-1+\frac{1+z}{2h(z)^2}\frac{d}{dz}h(z)^2
\end{equation}
Therefore, $(dh(z)^{2}/dz)_{z=0}=2(1+q_{0})$($q_{0}$ is the present
deceleration parameter). Evaluating $(dh(z)^{2}/dz)_{z=0}$ from
Eq.(\ref{eq:H2(m=3n)}) and solving with respect to $\eta$, we get:
\begin{equation}                        \label{eq:parameter term}
\eta=\frac{2q_{0}+ 2W_{\Lambda,0}-W_{m,0}-W_{X,0}}{3W_{X,0}}
\end{equation}
So far, the physical meaning of parameter term $\eta$ has clearly
stood before us. It reflects the combination of density parameters
and deceleration parameter at present. In other words, if we know
the values of density parameters and deceleration parameter, we
would know the ranges of $w_{0}$ and $a^{w}$. In addition, from
Eq.(\ref{eq:parameter term}), we also know:
$2(q_{0}+W_{\Lambda,0})>W_{m,0}+W_{X,0}$.

\subsection{The $m\neq3\alpha$ case}                     \label{sec:m><3n}
In this section, we have $m\neq3\alpha$ and Eq.(\ref{eq:pp}) has the
following result:
\begin{equation}        \label{eq:rho3}
\rho_{X}^{\alpha}=\rho_{X,0}^{\alpha}[(1-\frac{3\alpha}{3\alpha-m}\eta)x^{3\alpha}+\frac{3\alpha}{3\alpha-m}\eta
x^{m}]
\end{equation}
where $x$ is defined by $x=a_{0}/a$ called reduced scale factor. We
can see that dynamically changing part dark energy density evolves
depending on two terms of $x^{3\alpha}$ and $x^{m}$, and the values
of pure parameters term $\frac{3\alpha}{3\alpha-m}\eta$ determine
the effect. Additionally, the equation
(\ref{eq:rho3}) also tells us that $\rho_{X}^{\alpha}$ decreases
while the scale factor $a$ increases.

Combined with the EOS $p_{DE}=w\rho_{DE}$, the state parameter $w$
is expressed as:
\begin{equation}
w=-\frac{\frac{A_{0}}{a^{m}\rho_{X,0}^{\alpha}}\rho_{X}+\rho_{\Lambda}}{\rho_{X}+\rho_{\Lambda}}
\end{equation}
. Taking (\ref{eq:rho3}) into consideration, we can get the
following relation:
\begin{displaymath}
x^{3\alpha-m}\left( \begin{array}{l}
<\\
=\\
>
\end{array} \right)
-\frac{m}{(3\alpha-m)/\eta-3\alpha}
\end{displaymath}
\begin{displaymath}
provided
\left(
\begin{array}{l}
w<-1\\
w=-1\\
w>-1
\end{array} \right)
\end{displaymath}
Through the evolution of scale factor $a$, the value of
$x^{3\alpha-m}$ would satisfy the above relation, that is, the
$w=-1$ crossing phenomenon would occur. The different ranges of the
present state parameter $w_{0}$ and the scale factor $a^{w}$
corresponding to the state barrier($w=-1$) can be shown from the
following table no matter whether $3\alpha>m$ or $3\alpha<m$.

\begin{tabular*}{7cm}{@{\extracolsep {\fill}}c|c|c}
 \multicolumn{3}{l}{Table \ref{table:m!=3n}.  Parameter ranges and $w=-1$  }\\
  \hline
  $\eta$ & $w_{0}$& $a^{w}(w=-1)$ \\
  \hline
  $=1$ &$ =-1$ & $=a_{0}$\\
  \hline
  $<1$ & $>-1$ & $<a_{0}$ \\
  \hline
  $>1$ & $<-1$ & $>a_{0}$\\
  \hline
\end{tabular*}  \label{table:m!=3n}

Note: it is similar to table \ref{table:m=3n}, but the Judge $\eta$
has changed.

Corresponding to Eq.(\ref{eq:H2(m=3n)}), we can write out the
expression of the $m\neq3\alpha$ case as:
\begin{eqnarray}
h^{2}&=&W_{m,0}(1+z)^{3}+W_{X,0}(1+z)^{3} [1+\nonumber\\
      &&+\frac{3\alpha}{3\alpha-m}\eta((1+z)^{m-3\alpha}-1)]
      ^{\frac{1}{\alpha}}\nonumber\\
      &&+W_{\Lambda,0}+W_{k,0}(1+z)^{2}
\end{eqnarray}
It would be interesting to discuss the behavior around $a=a_{0}$.
Thus, we can expand the
$[1+\frac{3\alpha}{3\alpha-m}\eta((1+z)^{m-3\alpha}-1)]
      ^{\frac{1}{\alpha}}$ around $z=0$, preserve the first order of $z$, and get:
\begin{eqnarray}                            \label{eq:H20(m><3n)}
h^{2}&=&\widetilde{W}_{m,0}(1+z)^{3}-3W_{X,0}\eta(1+z)^{4}+\nonumber\\
      &&+W_{\Lambda,0}+W_{k,0}(1+z)^{2}
\end{eqnarray}
where $\widetilde{W}_{m,0}$ is defined by
\begin{equation}
\widetilde{W}_{m,0}=W_{m,0}+W_{X,0}(1+3\eta)
\end{equation}
The effective matter component also gets the contribution from the
dark energy component.

With the use of Eq.(\ref{eq:q}), we can get the Judge as:
\begin{equation}                        \label{eq:parameter term2}
\eta=-\frac{2q_{0}+ 2W_{\Lambda,0}-W_{m,0}-W_{X,0}}{3W_{X,0}}
\end{equation}
It results in:$2(q_{0}+W_{\Lambda,0})<W_{m,0}+W_{X,0}$ which is
quite different from $m=3\alpha$ case, and this difference can be
used as to distinguishing the two cases.

\section{Remarks on $A(x)$}                  \label{sec:The Taylor expansion case}
In the above treatment a$A(a)=A_{0}a^{-m}$ as in the last section is just one of the
possible forms for $A(a)$. Generally, we are not able to obtain $A(a)$ directly
from theoretical analysis. However, as to the far future universe in
which $x$(here we also define $x=a_{0}/a$) will approach to zero with
redshift negative if we define today's redshift value as zero,
the behaviour of $A(a)$ is then determined by the small quantity
$x$. It is natural to think of Taylor series expansion. We
substitute $A(a)$ by $A(x)$ for simplification,  and then expand $A(x)$.
\begin{eqnarray*}
    A(x)&=&\sum_{m}\frac{A^{(m)}(0)}{m!}x^{m}\\
    &=&A(0)+A'(0)x+\frac{1}{2}A''(0)x^{2}+\nonumber\\
    &&+\cdots+\frac{1}{m!}A^{(m)}(0)x^{m}+\cdots \label{eq:expansion A}
\end{eqnarray*}
Taking it into the expression of EOS, the pressure becomes
\begin{equation}\label{eq:expansion p}
p_{X}=\sum_{m}\frac{A_{(0)}^{(m)}x^{m}}{m!\rho_{X}^{\alpha}}\rho_{X}
\end{equation}
Compared with $A(a)=A_{0}a^{-m}$, the above expression is a
combination of Eq.(\ref{eq:state equation}) with different values of $m$.
This property may also imply that at the different cosmic stages
there is an accordant effective value of $m$ in the EOS
(ref{eq:state equation}) to describe the dark energy fluid. As for
far future stage, $m$ takes zero after neglecting small quantities.

Calculating density from Eq.(\ref{eq:pp}), we get
\begin{equation*}
    \rho_{X}^{\alpha}=\sum_{m}\frac{1}{m!}\frac{3\alpha}{3\alpha-m}A_{(0)}^{(m)}x^{m}
\end{equation*}

Often, $\rho_{X}$ and $a$ depend on each other and we can
deduce out the interrelationship of them from integration
formula(\ref{eq:pp}). Thus, pressure $p_{X}$ can be at last
expressed as a function with one variable , for example,
$p_{X}(\rho_{X})$, but not the two variables function
$p_{X}(\rho_{X},a)$. As for EOS
$p_{X}=-\frac{A(x)}{\rho_{X}^{\alpha}}\rho_{X}$, $A(x)$ and $\alpha$
are both used to determine $p_{X}(\rho_{X})$ and at last
$p_{X}(\rho_{X})$ has no relation with $\alpha$. According
to this assumpation, we would try to choose $A(x)$ to satisfy this requirment. Through a series of analysis, we finally get
\begin{equation*}
    A(x)=B^{\alpha}(1+x)^{3\alpha-1}
\end{equation*}
where $B$ is a lower limit for density $\rho_{X}^{\alpha}$ as
following. Thus, $\rho_{x}$ becomes
\begin{equation*}
    \rho_{x}=B(1+C^{1}_{3\alpha}+\cdots+C^{m}_{3\alpha}x^{m}+\cdots+x^{3\alpha})^{a/\alpha}=B(1+x)^{3}
\end{equation*}
where $C^{m}_{3\alpha}=\frac{(3\alpha)!}{m!(3\alpha-m)!}$ and $B$ is
the density lower limit defined by \[\lim_{x \rightarrow
0}\rho_{X}=B\] As a result, the EOS
\begin{equation}\label{eq:EOS}
p_{X}=-B^{\frac{1}{3}}\rho_{X}^{\frac{2}{3}}
\end{equation}
 is irrelevant
to $\alpha$. Then, we will discuss what information about universe
 this EOS gives.

\subsection{The expansion of universe}      \label{sec:The expansion of universe}                          
 For the purpose to show the extended Chaplygin gas contribution, we can temporally neglect
$\rho_{m}$ and $\rho_{\Lambda}$ in Eq.(\ref{eq:Fri}) and assume the
universe is flat for simplification. Solve Eq.(\ref{eq:Fri}) and get
\begin{equation}
\ln\frac{2\sqrt{1+x}+x+2}{(3+2\sqrt{2})x}-2\sqrt{\frac{1}{x+1}}+\sqrt{2}=(t-t_{0})B^{\frac{1}{2}}
\end{equation}
\begin{figure}
\begin{center}
\includegraphics{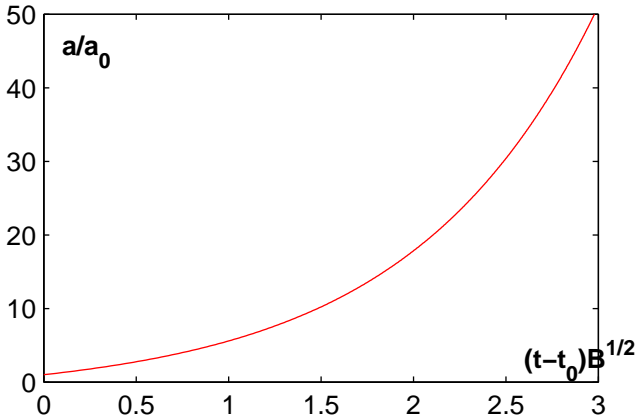}
\caption{It shows us the relationship of $a(t)$ and $t$. We assume
parameters $a_{0}=1$, $t_{0}=0$, $B=1$.}
 \label{fig:at}
\end{center}
\end{figure}
where $a_{0}$ and $t_{0}$ is the present scale factor and age of
universe, respectively. This function can easily be shown in  figure
\ref{fig:at}, on which the reduced scale factor increases
accompanying the elapse of cosmic time. And also $\ddot{a}>0$
implies our universe at the acceleration phase of state.

\subsection{ contributions from dynamical dark energy}                            
Under the help of Eq.(\ref{eq:Fri}), reduced Hubble parameter $h(z)$
can take the following form:
\begin{eqnarray}                \label{eq:matter}
h(z)^{2}&=&W_{m,0}(1+z)^{3}+W_{k,0}(1+z)^2+\nonumber\\
&&+W_{X,0}\frac{B}{\rho_{X,0}}(2+z)^{3}+W_{\Lambda,0}
\end{eqnarray}
As for `$X$-term' in above formula, it is easy to deploy $(2+z)^3$
to get $1+z$, $(1+z)^2$, $(1+z)^3$ and constant terms. Then combined
them with the other terms, we can get:
\begin{eqnarray}\label{eq:extent W}
h(z)^{2}&=&\widetilde{W}_{m,0}(1+z)^{3}+\widetilde{W}_{k,0}(1+z)^2+\nonumber\\
&&+W_{X,0}\frac{3B}{\rho_{X,0}}(1+z)+\widetilde{W}_{\Lambda,0}
\end{eqnarray}
where
\begin{eqnarray*}
    \widetilde{W}_{m,0}&=&W_{m,0}+W_{X,0}\frac{B}{\rho_{X,0}},\quad
    \widetilde{W}_{k,0}=W_{k,0}+W_{X,0}\frac{3B}{\rho_{X,0}}
    \nonumber\\
    &&\widetilde{W}_{\Lambda,0}=W_{\Lambda,0}+W_{X,0}\frac{B}{\rho_{X,0}}
\end{eqnarray*}
$\widetilde{W}_{m,0}$, $\widetilde{W}_{k,0}$,and
$\widetilde{W}_{\Lambda,0}$ are the effective density
parameters and show the $X$-term  behaviors. Noting our
prerequisite at cosmic far future evolution stage, the reduced Hubble parameter can be
approximately written as
\begin{equation*}
    h(z)^{2}\simeq\widetilde{W}_{\Lambda,0}
\end{equation*}
which means the de sitter phase. If our prerequisite is not so strict,
we can see that for the bigger values of $x$, $h(z)^{2}\simeq
\widetilde{W}_{m,0}(1+z)^{3}$ that means a dust-like phase. On the whole ,
 our universe has experienced a
phase transition from dust-like stage to a de sitter stage during the
process of cosmic evolution.

\subsection{The $w=-1$ crossing problem}
In section two we have discussed the crossing problem in
general cases, and now we reconsider it with the specific situation.
Taking EOS into the state parameter formula
$w=\frac{p_{DE}}{\rho_{DE}}$, we get:
\begin{equation}
w=-\frac{(\frac{B}{\rho_{X}})^{\frac{1}{3}}\rho_{X}+
\rho_{\Lambda}}{\rho_{X}+\rho_{\Lambda}}
\end{equation}
It is obvious that $w>-1$, that is to say, there is no crossing
phenomenon in this specific model and this particular case belongs to
a quintessence-like dark energy.

\section{Conclusion and discussion} \label{sec:Conclusion}
In this paper, in order to discuss the properties of dark energy with
more possibilities, which is
used to explain currently cosmic accelerating expansion, we have extended the
\emph{Chaplygin gas} model by which the Dark Fluid concept can be realized with its
contributions to effective universe compositions. Through a series of
mathematical treatment, it shows that there are indeed some cases
that can realize state parameter $w=-1$ crossing.  Moreover, the
 physical meanings behind the parameter terms like
$3A_{0}/a_{0}^{3\alpha}\rho_{X,0}^{\alpha}$ in Sec.\ref{sec:m=3n}
are interpreted.

Firstly, as for the $m=3\alpha$ case, we find that $\rho_{X}$ first
increases a maximum, and then decreases
approaching to a constant near $0$. During the procedure, the pressure $p$
 decreases  to $-\infty$. At the neighborhood
of $a=a_{0}$, the behavior of the density $\rho_{X}$ is similar to
non-relative matter with $\rho\propto a^{3}$. With further analysis,
we find that the state parameter $w$ can cross its traditional
barrier $w=-1$ smoothly without singularity. By the
calculations of the reduced Hubble parameter $h(z)$
(\ref{eq:H20(m=3n)}), we find that the dynamically
changing part of dark energy density is effectively equivalent to the
sum of the matter-like part and radiation-like part. After
considering the deceleration parameter $q(z)$, the parameter term
$\frac{3A_{0}}{a_{0}^{3\alpha}\rho_{X,0}^{\alpha}}$
 is explicitly expressed as combination of the present density parameters
$W_{i,0}$(subscript $i=m,X,\Lambda,k$) and the present deceleration
parameter $q_{0}$.

Secondly, as to the $m\neq3\alpha$ case, mostly it is similar to the
$m=3\alpha$ case. However,  different from the former case, the density
$\rho_{X}^{\alpha}$ is made up of $(a_{0}/a)^{3\alpha}$-term and
$(a_{0}/a)^{m}$-term,  and it is equivalent to total effects of both a
non-relative matter and some kinds of dark energy fluid with density
$\rho_{X}=(a_{0}/a)^{m}$. Obviously, from Eq.(\ref{eq:rho3})
$\rho_{X}$ always decreases to approach zero with the scale factor
$a(t)$ increasing.

Thirdly, after describing the general situations, we also discuss a
set of specific values of $m=3\alpha-1=0$.  Then, with parameter
independent assumpation,
concrete calculations show that the universe is in the state of
accelerating expansion with $\ddot{a}>0$. At last, we compare our
model with the typical state parameter $w=p_{DE}/\rho_{DE}$
obtaining $w>-1$, which means that there is never $w=-1$ crossing
phenomenon in this case. By discussing the reduced Hubble
parameter $h(z)$, we know that the dark energy fluid is evolving
equivalently to a fluid with
$\rho_{X}=3B(\frac{1}{a_{0}}+\frac{1}{a})$.

To sum up, we have discussed the extended
\emph{Chaplygin gas} model in detail, with the hope to show more possible
properties for dark energy and to realize the concept of
Dark Fluid
for a suggestive describing the cosmic dark components. To
constrain the EOS characterizing parameters, work will be done to maximize the
following likelihood function(see reference \cite{SC}):$L\propto
exp[-\chi^{2}(P)/2]$. And also in this paper, interactions between
the different components of cosmology are not considered. We will publish
the contents elsewhere soon\cite{mhr}.
\section{Acknowledgement}

We thank Prof.S.D.Odintsov  for reading the manuscript with
helpful comments and Profs. I. Brevik and Lewis H.Ryder for lots
of interesting discussions. This work is partly supported by NSF
and Doctoral Foundation of China.

\end{document}